# Integer and fractional Chern insulators in twisted bilayer MoTe$_2$

Yihang Zeng[1*], Zhengchao Xia[2*], Kaifei Kang[2], Jiacheng Zhu[2], Patrick Knüppel[2], Chirag Vaswani[2], Kenji Watanabe[3], Takashi Taniguchi[3], Kin Fai Mak[1,2,4**], Jie Shan[1,2,4**]

[1]Department of Physics, Cornell University, Ithaca, NY, USA.
[2]School of Applied and Engineering Physics, Cornell University, Ithaca, NY, USA.
[3]National Institute for Materials Science, Tsukuba, Japan.
[4]Kavli Institute at Cornell for Nanoscale Science, Ithaca, NY, USA.

*These authors contributed equally.
**Email: kinfai.mak@cornell.edu; jie.shan@cornell.edu

**Chern insulators, which are the lattice analogs of the quantum Hall states, can potentially manifest high-temperature topological orders at zero magnetic field to enable next-generation topological quantum devices [1-4]. To date, integer Chern insulators have been experimentally demonstrated in several systems at zero magnetic field [3, 5-11], but fractional Chern insulators have been reported only in graphene-based systems under a finite magnetic field [12, 13]. The emergence of semiconductor moiré materials [14, 15], which support tunable topological flat bands [16, 17], opens a new opportunity to realize fractional Chern insulators [18-20]. Here, we report the observation of both integer and fractional Chern insulators at zero magnetic field in small-angle twisted bilayer MoTe$_2$ by combining the local electronic compressibility and magneto-optical measurements. At hole filling factor $\nu = 1$ and 2/3, the system is incompressible and spontaneously breaks time reversal symmetry. We determine the Chern number to be 1 and 2/3 for the $\nu = 1$ and $\nu = 2/3$ gaps, respectively, from their dispersion in filling factor with applied magnetic field using the Streda formula. We further demonstrate electric-field-tuned topological phase transitions involving the Chern insulators. Our findings pave the way for demonstration of quantized fractional Hall conductance and anyonic excitation and braiding [21] in semiconductor moiré materials.**

## Main

Fractional Chern insulators (FCIs), which can in principle host the fractional quantum Hall effect and non-Abelian excitations at zero magnetic field, are highly sought-after phases of matter in condensed matter physics [22-28]. The experimental realization of FCIs may also revolutionize other fields, such as topological quantum computation [21]. But FCIs have proven notoriously challenging to realize experimentally because they require not only a topological flat band but also particular quantum band geometry [17-19, 29-33]. Band-structure engineering by forming moiré superlattices has emerged as a powerful approach to realize topological flat bands [11, 14, 15, 29, 34, 35]. A recent experiment has shown that FCIs can be stabilized in magic-angle twisted bilayer graphene at about 5 T, where the magnetic field is mainly responsible for redistributing the Berry curvature of the original topological bands [13]. With widely tunable electronic properties, moiré materials based on transition metal dichalcogenide (TMD) semiconductors have been predicted to support topological flat bands with appropriate band geometry to favor FCIs at zero magnetic field [18-20].

Of particular interest are small-angle twisted TMD homobilayers of the AA-stacking type (Fig. 1a). They support a honeycomb moiré lattice with two sublattices residing in two different layers [16, 17]. The topmost moiré valence bands are composed of the spin-valley locked states from the K or K' valley of the monolayers. Theoretical studies have shown that the complex interlayer hopping between the sublattice sites can induce topological moiré valence bands with non-zero spin/valley-resolved Chern numbers ($C$) [16, 17], and for certain twist angles, the topmost moiré band (with $|C| = 1$) is nearly flat and exhibits a flat Berry curvature distribution [16-19]. This opens the possibility of stabilizing FCIs at fractional fillings. Here we report the observation of an integer Chern insulator (CI) at $\nu = 1$ and FCI at $\nu = 2/3$ under zero magnetic field in 3.4-degree twisted bilayer MoTe$_2$ (tMoTe$_2$). The filling factor $\nu$ measures the hole doping density ($n$) in units of the moiré unit cell density ($n_M$), and $\nu = 1$ corresponds to quarter-band filling. These states display hallmarks of a CI. Specifically, they are incompressible, spontaneously break time reversal symmetry (TRS), linearly disperse in doping density with applied magnetic field, and carry an orbital magnetization that jumps across the charge gap. Furthermore, as the interlayer potential difference increases, our experiment at 1.6 K indicates a continuous topological phase transition from the integer CI to a topologically trivial Mott insulator, whereas the FCI becomes compressible.

**Correlated insulators with spontaneously broken TRS**
To search for the CIs, we perform local electronic compressibility measurements on dual-gated devices of tMoTe$_2$, in which the hole doping density ($n$) and the out-of-plane electric field ($E$ or the interlayer potential difference) can be independently controlled. To access the embedded sample, we apply a recently developed optical readout method for the chemical potential [36]. Figure 1a illustrates the device schematic. A high-quality WSe$_2$ monolayer is inserted between the sample and the top gate electrode as a sensor, and is capacitively coupled to the sample. The doping density in the sample is determined from the bias voltages and the geometrical capacitances of the device, which are calibrated locally based on the optically detected quantum oscillations in the sensor (Methods and Extended Data Fig. 1). We calibrate the moiré density ($n_M = (3.2 \pm 0.2) \times 10^{12}$ cm$^{-2}$, corresponding to twist angle $3.4 \pm 0.1$ degrees) using the density difference between the pronounced insulating states at $\nu = 1$ and $\nu = 2$. The lattice reconstructions are not significant in this twist angle range and relatively uniform moiré lattices can be obtained. The spatial resolution of the measurements is diffraction-limited to about 1 μm. Unless otherwise specified, the measurements are performed at 1.6 K. Details are provided in Methods.

Figure 1b (top panel) shows the doping dependence of the chemical potential $\mu(\nu)$ of tMoTe$_2$ with interlayer potential difference close to zero. The chemical potential is set to zero at its maximum value. Included is also its numerical derivative with respect to doping density (bottom panel), $\frac{d\mu}{d\nu}$, which is proportional to the electronic incompressibility (i.e. the inverse compressibility). As hole density increases, we observe a generally decreasing

$\mu$ or negative $\frac{d\mu}{d\nu}$ due to the strong electron-electron interaction in the flat band. On this background, there are three clearly observable chemical potential steps or incompressibility peaks at $\nu = 2$, $\nu = 1$ and $\nu = 2/3$. The insulating states at fractional band fillings ($\nu = 1$ and $\nu = 2/3$) are correlated in nature. From the chemical potential step size, we determine the charge gap to be about 6 mV and 0.6 mV for the $\nu = 1$ and $\nu = 2/3$ states, respectively (Methods and Extended Data Fig. 2).

Furthermore, the two insulating states spontaneously break TRS. This is probed locally by magnetic circular dichroism (MCD) near the MoTe$_2$ exciton resonance energy on the same sample location (see Methods). The MCD here measures the sample out-of-plane magnetization [15]. Figure 1c illustrates MCD as a function of perpendicular magnetic field ($B$) for $\nu = 1$ (left panel) and $\nu = 2/3$ (right panel), respectively. Both states exhibit spontaneous MCD and a magnetic hysteresis with coercive field about 20 mT. This result is consistent with an earlier report for a 3.9-degree tMoTe$_2$ (Ref. [37]).

**Determination of the Chern number**
To verify whether these states are CIs, we measure the incompressibility as a function of doping density and perpendicular magnetic field up to 8 T. Figure 2a shows that both incompressible states disperse linearly towards larger filling factor with applied magnetic field. The slope, $\frac{d\nu}{dB}$, of the $\nu = 2/3$ state is 2/3 of the slope of the $\nu = 1$ state. For comparison, Fig. 2b shows the same measurement under a large interlayer potential difference ($E = -14$ mV/nm at $\nu = 1$), for which spontaneous MCD is not observed at $\nu = 1$ (Fig. 3b). The $\nu = 2/3$ state becomes compressible. The $\nu = 1$ state remains incompressible but does not disperse with magnetic field (the small deviation observed above 6 T is mainly from the relative sample-beam drift on the sub-micron level and the twist angle/moiré density disorder). As we discuss below, the incompressible state at $\nu = 1$ under a large interlayer potential difference is compatible with a topologically trivial Mott insulator when the charges are transferred to a single MoTe$_2$ layer and the problem becomes one with half-band filling in a triangular lattice [15].

The linear dispersion illustrated in Fig. 2a is a hallmark of a CI. We determine the Chern number for the $\nu = 1$ and $\nu = 2/3$ gaps from their slope using the Streda formula, $C = \frac{h}{e}\frac{d\nu}{dB}n_M$, with $h$ and $e$ denoting the Planck's constant and the elementary charge, respectively. The Streda formula relates one magnetic flux quantum that is added to the sample to pumping $C$ holes between the Chern bands [38]. We extract the center-of-mass of the incompressibility peak (Methods) for the $\nu = 1$ and $\nu = 2/3$ states at each magnetic field (empty symbols, Fig. 2c). The filled symbols are the corrected center positions after taking into account the small sample-beam drift (under high magnetic fields). We use the center-of-mass of the incompressibility peak of the Mott insulator (Fig. 2b) to calibrate the filling factor for each magnetic field. We find the Chern number for the $\nu = 1$ and $\nu = 2/3$ gaps to be $C = 1.0 \pm 0.1$ and $C = 0.63 \pm 0.08$, respectively, from the linear fits to the corrected center positions (see Methods for an independent calibration of $C$'s). We conclude that within the experimental uncertainty, the $\nu = 1$ state is an integer CI with Chern number 1 and the $\nu = 2/3$ state is a FCI with Chern number 2/3.

The CIs also possess an orbital magnetization ($M$) because of the presence of topological edge states [38, 39]. We can evaluate $M$ from the magnetic-field dependent chemical potential via the Maxwell's relation, $-\left(\frac{\partial \mu}{\partial B}\right)_\nu = \frac{1}{n_M}\left(\frac{\partial M}{\partial \nu}\right)_B$. Figure 2d shows $-\left(\frac{\partial \mu}{\partial B}\right)_\nu$ computed from the difference between the measured chemical potential at $B = 0$ T and 3 T (Extended Data Fig. 3). A peak is observed for both CIs. The integrated peak area (shaded) provides an estimate of the orbital magnetization change across the CI gap per moiré unit cell, $\frac{\Delta M}{n_M}$ (Ref. [39]). We estimate $\frac{\Delta M}{n_M} \approx 0.4 \, \mu_B$ and $0.05 \, \mu_B$ for the $\nu = 1$ and $\nu = 2/3$ state, respectively, where $\mu_B$ denotes the Bohr magneton. On the other hand, the orbital magnetization jump is determined by the gap size ($\Delta \mu$) as $\Delta M = C \frac{e}{h} \Delta \mu$, or equivalently, $\frac{\Delta M}{n_M} = C \mu_B \left(\frac{\Delta \mu}{W}\right)$ is given by the ratio of the gap size and the characteristic moiré bandwidth, $W = \frac{n_M h^2}{4\pi m}$ ($m$ denoting the electron mass) (Ref. [39]). Using the measured gap size, we estimate $\frac{\Delta M}{n_M} \approx 0.8 \, \mu_B$ and $0.06 \, \mu_B$ for the $\nu = 1$ and $\nu = 2/3$ CI states, respectively, which are comparable with the estimates from the measurement of $\left(\frac{\partial \mu}{\partial B}\right)_\nu$ (Fig. 2d). The moiré bandwidth $W \approx 8$ meV is comparable to recent theoretical calculations [40-42]. Remarkably, the $\nu = 1$ and $\nu = 2/3$ gap sizes are about 80% and 8% of the moiré bandwidth, respectively. The relatively large gap sizes highlight the strong electronic correlations in tMoTe$_2$.

**Topological phase transitions**
To map out the phase diagram of the CIs and study the topological phase transitions, we measure the incompressibility (Fig. 3a) and MCD (Fig. 3b) as a function of doping density and perpendicular electric field. A small magnetic field (20 mT) is applied to suppress the MCD fluctuations (likely from the magnetic domains). To correlate with the electrostatics phase diagram, we also measure the sample reflectance at the fundamental intralayer exciton resonance of MoTe$_2$ (Fig. 3c). (See Extended Data Fig. 4 for the reflectance and MCD spectra.) Strong reflectance signifies that at least one of the MoTe$_2$ layers is charge neutral because doping efficiently quenches the intralayer exciton resonance in TMDs [43]. We identify two distinct regions separated by the dashed lines. In the middle region (quenched reflectance), layers are hybridized, and charges are shared between two layers. Outside this region (strong reflectance), all the charges reside in one of the layers. Similar electrostatics phase diagrams have been reported in twisted bilayer WSe$_2$ and other related TMD moiré heterostructures [44-46]. Zero interlayer potential difference is shifted from $E_0 = 0$ to - 90 mV/nm in this device due to the presence of a built-in electric field (from the asymmetric device structure with a sensor layer).

We overlay the boundary of the charge-sharing region (dashed lines) on the incompressibility and spontaneous MCD maps. The $\nu = 1$ state appears incompressible throughout the phase diagram with weakened incompressibility near the boundary. The $\nu = 2/3$ state is incompressible only in the layer-hybridized region. On the other hand, the spontaneous MCD is observed over a broad doping range in the layer-hybridized region (the fine structures of the map may come from magnetic domains, which require further studies). We identify MCD hot spots around $\nu = 1$ and $\nu = 2/3$, which signify the emergence of CIs. The dashed lines therefore also provide the phase boundary for CIs. Outside, the correlated insulator at $\nu = 1$ is a Mott insulator, and the compressible state at $\nu = 2/3$ is likely a correlated Fermi liquid.

We examine the electric-field-tuned topological phase transitions in more detail. Figure 4a and 4b are the electric-field dependent charge gap and spontaneous MCD at $\nu = 1$, respectively. The vertical dashed lines denote the phase boundary. The chemical potential step measured at 1.6 K decreases from ~ 6 mV to a minimum of ~ 5 mV as $|E - E_0|$ approaches the boundary, and rapidly increases with further increase of the electric field. The observed gap minimum suggests gap closing at the critical point that is broadened by finite temperature and/or disorder. Correlated, the MCD (1.6 K) decreases rapidly beyond the phase boundary. As temperature increases, the MCD decreases and the ferromagnetic phase space narrows continuously. We estimate the highest critical temperature to be $T_c \approx$ 13 K. Likewise, we show in Fig. 4c the chemical potential step that can be determined for the $\nu = 2/3$ state. It is ~ 0.6 mV in the FCI phase and vanishes beyond the phase boundary. The spontaneous MCD in Fig. 4d also vanishes beyond the phase boundary and decreases continuously with increasing temperature. The critical temperature is estimated to be $T_c \approx$ 5 K.

For the $\nu = 1$ phase transition, we observe two energy scales, one for charge localization from electron correlations and the other for onset of magnetic order from exchange interactions. The CI emerges after long-range magnetic order develops below $T_c$. The phenomenology is similar to that observed in graphene moiré systems [6-8, 11] and AB-stacked MoTe$_2$/WSe$_2$ moiré bilayers [10], where the energy scale for charge localization is generally higher than that for magnetism. Interestingly, the two energy scales for the $\nu = 2/3$ transition are comparable; the observation deserves further investigations. Furthermore, our experiment suggests that both topological phase transitions are continuous, which occur by closing the charge gap and are in agreement with recent mean-field calculations for TMD moiré materials [17, 47-49]. The transitions are distinct from the Chern-to-Mott insulator transition observed in AB-stacked MoTe$_2$/WSe$_2$ moiré bilayers, where no charge gap minimum is observed at the critical point to support broadened gap closing [10, 36].

**Conclusions**
We demonstrate an integer and a fractional CI at zero magnetic field in small-angle tMoTe$_2$ by the local measurements of the electronic compressibility and TRS breaking. We also observe evidence for a continuous topological phase transition for both CI states that is induced by the interlayer potential difference. Our findings leave open many questions, such as the nature of the FCI and possible existence of FCIs in tMoTe$_2$ and related materials that have no analogs in the fractional quantum Hall system. A pressing experimental task is to develop electrical contacts to these materials for transport measurements and for manipulation of the anyonic excitation for topological quantum applications. During the preparation of this manuscript, we learned about the work that reports signatures of FCIs in tMoTe$_2$ using optical spectroscopy techniques [50], as well as another work that reports integer CIs in tWSe2 using local compressibility measurements [51].

## Methods

### Device fabrication.

We fabricated the dual-gated devices by following the procedure reported elsewhere (e.g. Ref. [44, 52]). The device details for the chemical potential measurements are described in Ref. [36]. The main differences here are: both the sample and sensor are contacted by Pt electrodes, and the metal-semiconductor contacts are gated by additional contact gates, which are made of the standard few-layer graphite electrodes and hBN (hexagonal boron nitride) dielectrics. These steps improve the electrical contact to both the sample and sensor for the chemical potential measurements. A cross-sectional schematic and an optical micrograph of the device are shown in Extended Data Fig. 5. We calibrated the gate capacitances locally using the optically detected Landau level spectroscopy of the sensor (see below for details). We obtain the hBN thickness in the top gate ($d_{tg}$ = 6.1 nm), the bottom gate ($d_{bg}$ = 7 nm) and the sample-to-sensor gate ($d_s$ = 3.2 nm). The hBN thickness in the contact gates is approximately 15 nm.

### Optical measurements.

For all measurements, the tMoTe$_2$ device was mounted in a closed-cycle optical cryostat (Attocube, attoDRY2100). We focused broadband emission from light emitting diodes (LEDs) onto the device using a 40x low-temperature microscope objective (numerical aperture 0.8). The beam size is about 1 μm on the sample and the incident power is limited to < 0.8 nW to minimize the sample heating effect. We collected the reflected light from the sample using the same objective and sent it to a grating spectrometer equipped with a silicon charge-coupled device (CCD) and an InGaAs linear array photo-detector. Details on the reflectance contrast, MCD and chemical potential measurements have been discussed elsewhere (e.g. Ref. [36, 53]). Specifically, we used LEDs with spectral range of 700-760 nm and the Si CCD for chemical potential measurements on the monolayer WSe$_2$ sensor, and LEDs with spectral range of 1030-1130 nm and the InGaAs detector for the reflectance contrast and MCD measurements on tMoTe$_2$. The relative sample-beam drift is typically small for low magnetic fields. It is on the order of 0.5 – 1 μm for magnetic fields > 6 T.

### Determination of the filling factor of incompressible states.

To determine the filling factor of the incompressible states in Fig. 2a,b, we computed for each magnetic field the center-of-mass of the incompressibility peak over a filling factor range that nearly covers the entire peak. We also used a slightly larger window for $\nu$ = 1 than for $\nu = 2/3$. An example is shown in Extended Data Fig. 2. The slope of the magnetic-field dependent center-of-mass filling factor is used to determine the Chern number of the incompressible states by the Streda formula.

### Calibration of the moiré density.

Accurate calibration of the moiré density in tMoTe$_2$ is required to determine the Chern number of the incompressible states. We achieve this by studying the quantum oscillations in the sensor (Extended Data Fig. 1). We first heavily hole dope the sample with the gate biases and fix them. We then monitor the reflectance contrast spectrum of the sensor (monolayer WSe$_2$) as a function of bias between the sample and sensor, which continuously tunes the hole doping density in the sensor. Compared to the sample, the sensor has high carrier mobilities, and quantum oscillations can be readily observed in the optical conductivity under a constant magnetic field of 8.8 T. Using the voltage interval between the two-fold degenerate Landau levels at high hole doping densities [54], we determine the geometrical capacitance between the sample and sensor ($C_s \approx 8.3$ μFcm$^{-2}$). The bottom gate capacitance ($C_{bg}$) is determined by calibrating the capacitance lever arm (Extended Data Fig. 6). With the known capacitances, we determine the moiré density of the sample, $n_M = (3.2 \pm 0.2) \times 10^{12}$ cm$^{-2}$, using the voltage difference between the $\nu = 1$ and $\nu = 2$ incompressible states. This corresponds to a twist angle of $3.4 \pm 0.1$ degrees.

**Independent verification of the Chern numbers.**
We independently verified the Chern number of the CIs by studying the electric-field and doping dependent incompressibility at a fixed magnetic field of 8.8 T (Extended Data Fig. 7). Measurements at a fixed magnetic field remove the relative sample-beam drift. We observe both the CI and the non-topological MI states at $\nu = 1$. We determine the center-of-mass filling factor of both states and their difference (0.070 ± 0.006). This corresponds to Chern number $C = 1.1 \pm 0.1$ for the CI. Since the slope of the linearly dispersing $\nu = 2/3$ state is 2/3 of that of the $\nu = 1$ state (Fig. 2), the Chern number of the $\nu = 2/3$ state is $C = 0.70 \pm 0.06$.

**Estimate of the $\nu = 2/3$ chemical potential step.**
Compared to the $\nu = 1$ state, the $\nu = 2/3$ chemical potential step is substantially smaller, and it sits on a large negative compressibility background. Direct readout of the chemical potential step from $\mu(\nu)$ is therefore difficult. In our analysis, we first subtracted the constant negative incompressibility background from $\frac{d\mu}{d\nu}$ around the $\nu = 2/3$ state, and integrated the incompressibility peak area to obtain the chemical potential step. An example is shown in Extended Data Fig. 2. We performed the same analysis on the $\nu = 1$ state and obtained consistent values to that from direct readout of the chemical potential step.

**Acknowledgement**
We thank Liang Fu for fruitful discussions.

**References**
1. Haldane, F.D.M. Model for a Quantum Hall Effect without Landau Levels: Condensed-Matter Realization of the "Parity Anomaly". *Physical Review Letters* **61**, 2015-2018 (1988).
2. Hasan, M.Z. & Kane, C.L. Colloquium: Topological insulators. *Reviews of Modern Physics* **82**, 3045-3067 (2010).
3. Liu, C.-X., Zhang, S.-C. & Qi, X.-L. The Quantum Anomalous Hall Effect: Theory and Experiment. *Annual Review of Condensed Matter Physics* **7**, 301-321 (2016).
4. Chang, C.-Z., Liu, C.-X. & MacDonald, A.H. Colloquium: Quantum anomalous Hall effect. *Reviews of Modern Physics* **95**, 011002 (2023).

5. Chang, C.-Z. et al. Experimental Observation of the Quantum Anomalous Hall Effect in a Magnetic Topological Insulator. *Science* **340**, 167-170 (2013).
6. Sharpe, A.L. et al. Emergent ferromagnetism near three-quarters filling in twisted bilayer graphene. *Science* **365**, 605-608 (2019).
7. Serlin, M. et al. Intrinsic quantized anomalous Hall effect in a moiré heterostructure. *Science* **367**, 900-903 (2020).
8. Chen, G. et al. Tunable correlated Chern insulator and ferromagnetism in a moiré superlattice. *Nature* **579**, 56-61 (2020).
9. Deng, Y. et al. Quantum anomalous Hall effect in intrinsic magnetic topological insulator MnBi2Te4. *Science* **367**, 895-900 (2020).
10. Li, T. et al. Quantum anomalous Hall effect from intertwined moiré bands. *Nature* **600**, 641-646 (2021).
11. Liu, J. & Dai, X. Orbital magnetic states in moiré graphene systems. *Nature Reviews Physics* **3**, 367-382 (2021).
12. Spanton, E.M. et al. Observation of fractional Chern insulators in a van der Waals heterostructure. *Science* **360**, 62-66 (2018).
13. Xie, Y. et al. Fractional Chern insulators in magic-angle twisted bilayer graphene. *Nature* **600**, 439-443 (2021).
14. Kennes, D.M. et al. Moiré heterostructures as a condensed-matter quantum simulator. *Nature Physics* **17**, 155-163 (2021).
15. Mak, K.F. & Shan, J. Semiconductor moiré materials. *Nature Nanotechnology* **17**, 686-695 (2022).
16. Wu, F., Lovorn, T., Tutuc, E., Martin, I. & MacDonald, A.H. Topological Insulators in Twisted Transition Metal Dichalcogenide Homobilayers. *Physical Review Letters* **122**, 086402 (2019).
17. Devakul, T., Crépel, V., Zhang, Y. & Fu, L. Magic in twisted transition metal dichalcogenide bilayers. *Nature Communications* **12**, 6730 (2021).
18. Li, H., Kumar, U., Sun, K. & Lin, S.-Z. Spontaneous fractional Chern insulators in transition metal dichalcogenide moir\'e superlattices. *Physical Review Research* **3**, L032070 (2021).
19. Valentin Crépel, L.F. Anomalous Hall metal and fractional Chern insulator in twisted transition metal dichalcogenides. *arXiv:2207.08895* (2022).
20. Nicolás Morales-Durán, J.W., Gabriel R. Schleder, Mattia Angeli, Ziyan Zhu, Efthimios Kaxiras, Cécile Repellin, Jennifer Cano. Pressure--enhanced fractional Chern insulators in moiré transition metal dichalcogenides along a magic line. *arXiv:2304.06669* (2023).
21. Nayak, C., Simon, S.H., Stern, A., Freedman, M. & Das Sarma, S. Non-Abelian anyons and topological quantum computation. *Reviews of Modern Physics* **80**, 1083-1159 (2008).
22. Sheng, D.N., Gu, Z.-C., Sun, K. & Sheng, L. Fractional quantum Hall effect in the absence of Landau levels. *Nature Communications* **2**, 389 (2011).
23. Neupert, T., Santos, L., Chamon, C. & Mudry, C. Fractional Quantum Hall States at Zero Magnetic Field. *Physical Review Letters* **106**, 236804 (2011).
24. Tang, E., Mei, J.-W. & Wen, X.-G. High-Temperature Fractional Quantum Hall States. *Physical Review Letters* **106**, 236802 (2011).

25. Regnault, N. & Bernevig, B.A. Fractional Chern Insulator. *Physical Review X* **1**, 021014 (2011).
26. Qi, X.-L. Generic Wave-Function Description of Fractional Quantum Anomalous Hall States and Fractional Topological Insulators. *Physical Review Letters* **107**, 126803 (2011).
27. Wu, Y.-L., Bernevig, B.A. & Regnault, N. Zoology of fractional Chern insulators. *Physical Review B* **85**, 075116 (2012).
28. Parameswaran, S.A., Roy, R. & Sondhi, S.L. Fractional quantum Hall physics in topological flat bands. *Comptes Rendus Physique* **14**, 816-839 (2013).
29. Zhang, Y.-H., Mao, D., Cao, Y., Jarillo-Herrero, P. & Senthil, T. Nearly flat Chern bands in moir\'e superlattices. *Physical Review B* **99**, 075127 (2019).
30. Abouelkomsan, A., Liu, Z. & Bergholtz, E.J. Particle-Hole Duality, Emergent Fermi Liquids, and Fractional Chern Insulators in Moiré Flatbands. *Physical Review Letters* **124**, 106803 (2020).
31. Ledwith, P.J., Tarnopolsky, G., Khalaf, E. & Vishwanath, A. Fractional Chern insulator states in twisted bilayer graphene: An analytical approach. *Physical Review Research* **2**, 023237 (2020).
32. Repellin, C. & Senthil, T. Chern bands of twisted bilayer graphene: Fractional Chern insulators and spin phase transition. *Physical Review Research* **2**, 023238 (2020).
33. Wilhelm, P., Lang, T.C. & Läuchli, A.M. Interplay of fractional Chern insulator and charge density wave phases in twisted bilayer graphene. *Physical Review B* **103**, 125406 (2021).
34. Andrei, E.Y. & MacDonald, A.H. Graphene bilayers with a twist. *Nature Materials* **19**, 1265-1275 (2020).
35. Mai, P., Feldman, B.E. & Phillips, P.W. Topological Mott insulator at quarter filling in the interacting Haldane model. *Physical Review Research* **5**, 013162 (2023).
36. Xia, Z. et al. Optical readout of the chemical potential of two-dimensional electrons. *arXiv preprint arXiv:2304.09514* (2023).
37. Eric Anderson, F.-R.F., Jiaqi Cai, William Holtzmann, Takashi Taniguchi, Kenji Watanabe, Di Xiao, Wang Yao, Xiaodong Xu. Programming Correlated Magnetic States via Gate Controlled Moiré Geometry. *arXiv:2303.17038* (2023).
38. MacDonald, A.H. Introduction to the Physics of the Quantum Hall Regime. *arXiv:cond-mat/9410047* (1994).
39. Zhu, J., Su, J.-J. & MacDonald, A.H. Voltage-Controlled Magnetic Reversal in Orbital Chern Insulators. *Physical Review Letters* **125**, 227702 (2020).
40. Aidan P. Reddy, F.F.A., Yang Zhang, Trithep Devakul, Liang Fu. Fractional quantum anomalous Hall states in twisted bilayer MoTe2 and WSe2. *arXiv:2304.12261* (2023).
41. Chong Wang, X.-W.Z., Xiaoyu Liu, Yuchi He, Xiaodong Xu, Ying Ran, Ting Cao, Di Xiao. Fractional Chern Insulator in Twisted Bilayer MoTe2. *arXiv:2304.11864* (2023).
42. Wen-Xuan Qiu, Bohao Li, Xun-Jiang Luo, Fengcheng Wu. Interaction-driven topological phase diagram of twisted bilayer MoTe2. *arXiv:2305.01006* (2023).

43. Wang, G. et al. Colloquium: Excitons in atomically thin transition metal dichalcogenides. *Reviews of Modern Physics* **90**, 021001 (2018).
44. Xu, Y. et al. A tunable bilayer Hubbard model in twisted WSe2. *Nature Nanotechnology* **17**, 934-939 (2022).
45. Zhao, W. et al. Gate-tunable heavy fermions in a moiré Kondo lattice. *Nature*, 1-5 (2023).
46. Gu, J. et al. Dipolar excitonic insulator in a moiré lattice. *Nature Physics* **18**, 395-400 (2022).
47. Devakul, T. & Fu, L. Quantum Anomalous Hall Effect from Inverted Charge Transfer Gap. *Physical Review X* **12**, 021031 (2022).
48. Xie, Y.-M., Zhang, C.-P., Hu, J.-X., Mak, K.F. & Law, K.T. Valley-Polarized Quantum Anomalous Hall State in Moiré MoTe2/WSe2 Heterobilayers. *Physical Review Letters* **128**, 026402 (2022).
49. Pan, H., Xie, M., Wu, F. & Das Sarma, S. Topological Phases in AB-Stacked MoTe2/WSe2 Topological Insulators, Chern Insulators, and Topological Charge Density Waves. *Physical Review Letters* **129**, 056804 (2022).
50. Jiaqi Cai, E.A., Chong Wang, Xiaowei Zhang, Xiaoyu Liu, William Holtzmann, Yinong Zhang, Fengren Fan, Takashi Taniguchi, Kenji Watanabe, Ying Ran, Ting Cao, Liang Fu, Di Xiao, Wang Yao, Xiaodong Xu. Signatures of Fractional Quantum Anomalous Hall States in Twisted MoTe2 Bilayer. *arXiv:2304.08470* (2023).
51. Benjamin A. Foutty, C.R.K., Trithep Devakul, Aidan P. Reddy, Kenji Watanabe, Takashi Taniguchi, Liang Fu, Benjamin E. Feldman. Mapping twist-tuned multi-band topology in bilayer WSe2. *arXiv:2304.09808* (2023).
52. Wang, L. et al. One-Dimensional Electrical Contact to a Two-Dimensional Material. *Science* **342**, 614-617 (2013).
53. Tao, Z. et al. Valley-coherent quantum anomalous Hall state in AB-stacked MoTe2/WSe2 bilayers. *arXiv preprint arXiv:2208.07452* (2022).
54. Gustafsson, M.V. et al. Ambipolar Landau levels and strong band-selective carrier interactions in monolayer WSe2. *Nature Materials* **17**, 411-415 (2018).

# Figures

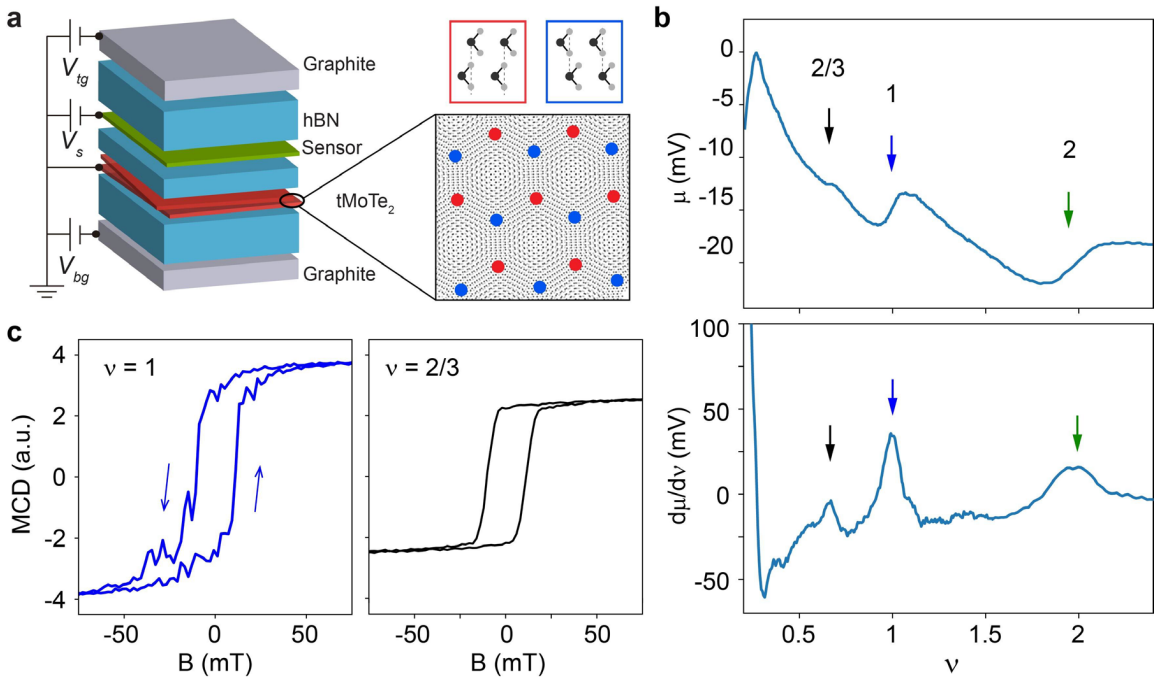

**Figure 1 | Ferromagnetic incompressible states in tMoTe$_2$. a,** Schematic of dual-gated devices of tMoTe$_2$ with a monolayer WSe$_2$ sensor for local electronic compressibility measurements. The sample is grounded, and $V_{tg}$, $V_s$ and $V_{bg}$ are the biases applied to the top graphite/hBN gate, the sensor, and the bottom graphite/hBN gate, respectively. Inset: tMoTe$_2$ forms a honeycomb moiré superlattice. Mo (Te) atoms in the top layer are aligned with Te (Mo) atoms in the bottom layer at the red (blue) sublattice sites. **b,** Chemical potential (top) and electronic incompressibility (bottom) as a function of $\nu$, the hole doping density in units of moiré density $n_M \approx 3.2 \times 10^{12}$ cm$^{-2}$. The chemical potential is set to zero at the maximum. Three prominent incompressible states at $\nu = 2/3$, 1 and 2 are marked. **c,** Magnetic-field dependence of the sample MCD at $\nu = 1$ (left) and $\nu = 2/3$ (right). Spontaneous MCD and magnetic hysteresis are observed. The MCD fluctuations at $\nu = 1$ likely reflects the presence of magnetic domains. In **b** and **c**, the interlayer potential difference is close to zero near $\nu = 1$ and $\nu = 2/3$ (with $V_s = 0.14$ V and $V_{bg}$ scanned).

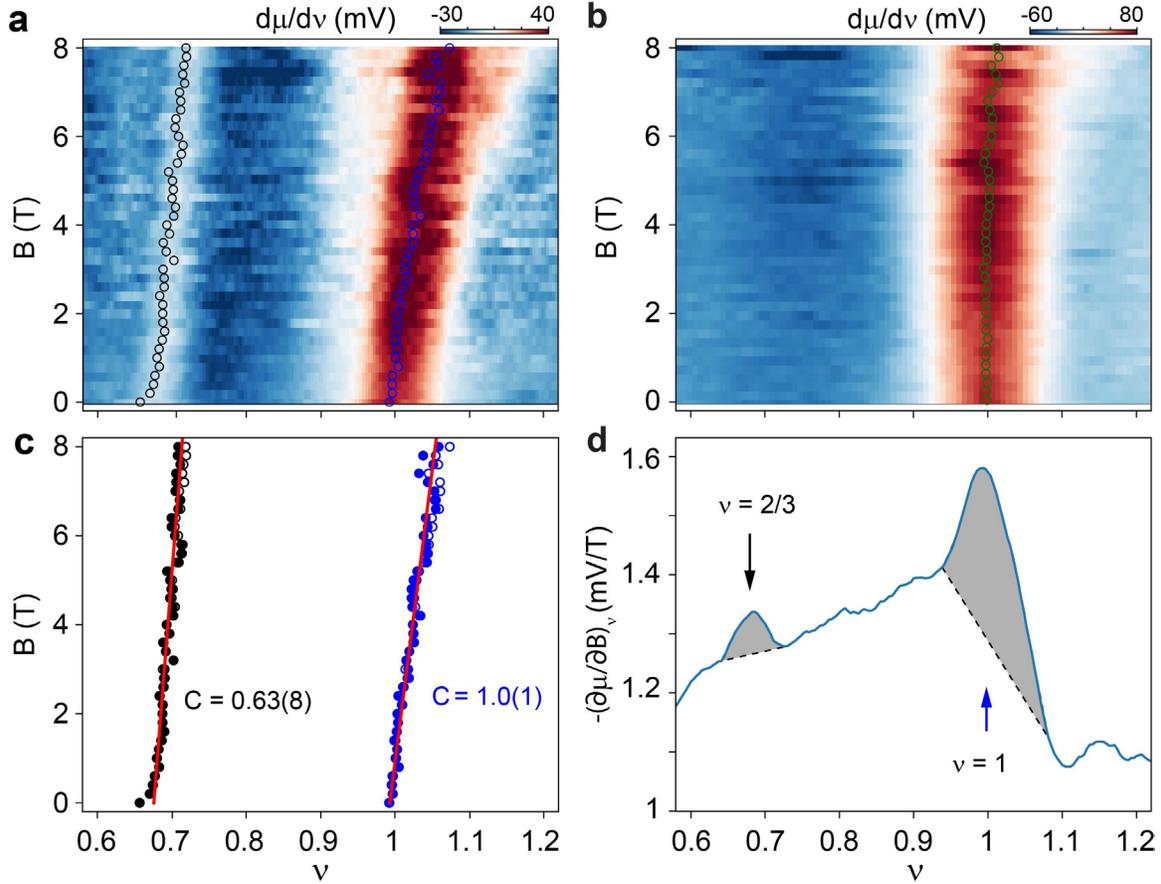

**Figure 2 | Integer and fractional Chern insulators. a,** Electronic incompressibility as a function of hole filling factor ($\nu$) and perpendicular magnetic field ($B$) near zero interlayer potential difference. Two linearly dispersing incompressible states are observed at $\nu = 1$ and $\nu = 2/3$. **b,** Same as **a** under a large interlayer potential difference. One non-dispersive incompressible state is observed at $\nu = 1$. Empty circles in **a, b** are the center-of-mass of the incompressibility peaks. A small sample-beam drift is present above ~ 6 T. **c,** Determination of the Chern number ($C$) of the $\nu = 1$ and $\nu = 2/3$ states. Empty circles: same as in **a**; filled circles: corrected incompressibility peak location using that of **b**; solid lines: linear fits to the filled circles. **d,** Filling dependence of $-\left(\frac{\partial \mu}{\partial B}\right)_\nu$ computed from the difference of $\mu(\nu)$ at 3 T and 0 T near zero interlayer potential difference. The peaks show the presence of in-gap orbital magnetization for the CIs. The peak areas (shaded) provide an estimate for the total orbital magnetization.

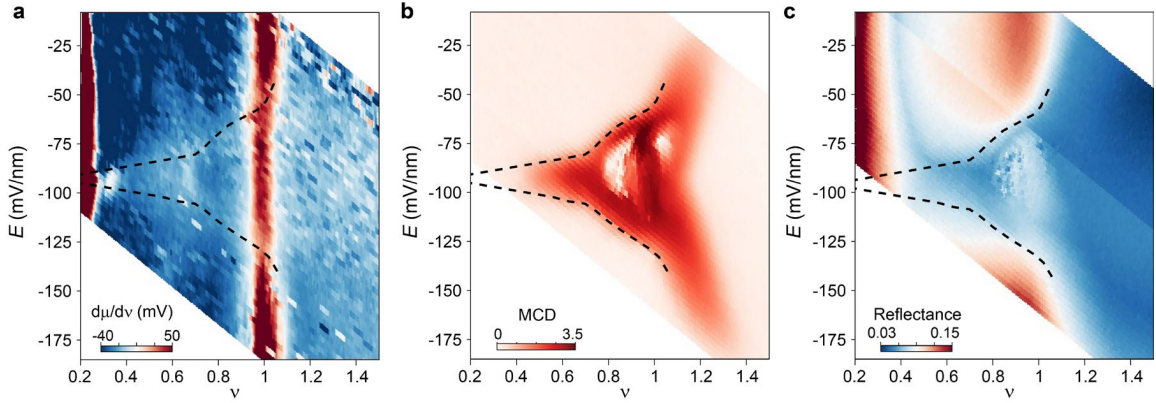

**Figure 3 | Phase diagram. a-c,** Electronic incompressibility (**a**), MCD (**b**) and optical reflectance at the intralayer exciton resonance (**c**) of tMoTe$_2$ as a function of hole filling factor ($\nu$) and perpendicular electric field ($E$). A small magnetic field (20 mT) is applied in **b** to reduce the MCD fluctuations (likely due to the presence of magnetic domains). The spontaneous MCD has a similar magnitude for the $\nu = 1$ and $\nu = 2/3$ states. In the layer-hybridized region between the dashed lines, the interlayer potential difference is small and charges are shared between the two layers.

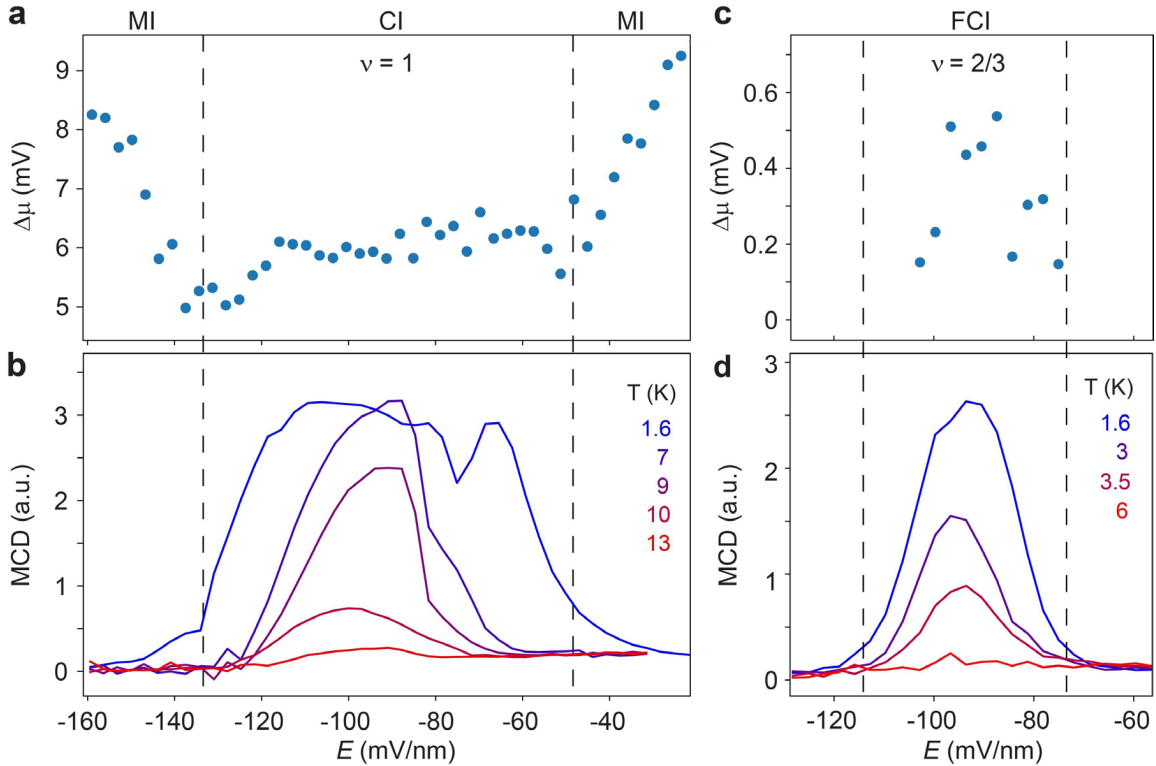

**Figure 4 | Topological phase transitions. a,b,** Electric-field dependence of the chemical potential step at 1.6 K (**a**) and the spontaneous MCD at representative temperatures (**b**) for the $\nu = 1$ state. The dashed lines (same as in Fig. 3) mark the boundary of the charge-sharing region. Chemical potential step minima are observed near the boundary. The spontaneous MCD vanishes beyond the boundary and above the critical temperature of about 13 K. The results indicate a continuous phase transition from a Chern insulator (CI) to a non-topological Mott insulator (MI). **c,d,** Same as **a,b,** for the $\nu = 2/3$ state. Both the chemical potential step and spontaneous MCD vanish beyond the boundary. The magnetic critical temperature is about 5 K.

**Extended Data Figures**

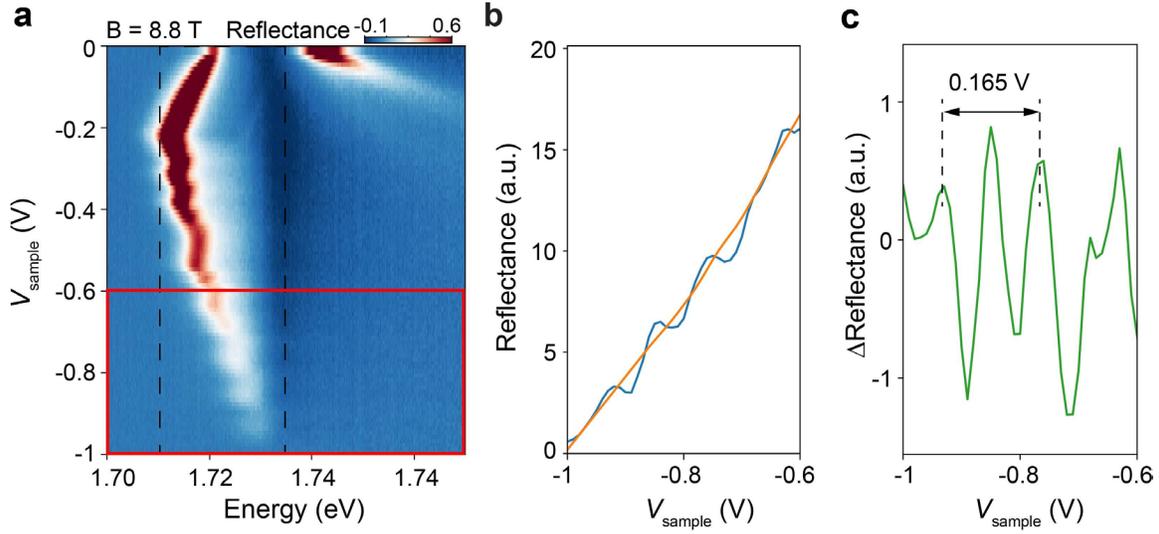

**Extended Data Figure 1 | Calibration of moiré density. a,** Dependence of the optical reflectance spectrum of the sensor on the sample-to-sensor bias voltage $V_{sample}$ (sensor grounded); the top and bottom gate voltages are kept constant; the sample is heavily hole-doped; the magnetic field is fixed at 8.8 T. Clear quantum oscillations in the attractive polaron energy and amplitude are observed with hole doping due to the formation of spin-valley-polarized Landau levels at high magnetic fields. **b,** Dependence of the integrated attractive polaron amplitude (over the spectral window bound by the dashed lines in **a**) on $V_s$ (blue curve). The orange curve represents the smooth background excluding the quantum oscillations. We only show data between -0.6 and -1 V, where the sensor is heavily hole-doped and the Landau levels are two-fold degenerate. **c,** Dependence of the oscillation amplitude (after removal of the smooth background) on $V_s$. The average distance between adjacent amplitude peaks at high hole doping densities is 82.5 mV; this corresponds to a change in the sensor doping density of $4.26 \times 10^{11}$ cm$^{-2}$ based on the known Landau level degeneracy (= 2) in this doping density range. The data allows us to accurately determine the sample-to-sensor geometrical capacitance $C_s \approx 8.3 \pm 0.2$ µFcm$^{-2}$, as well as the other gate capacitances (see Methods and Extended Data Fig. 6), from which a moiré density $n_M = (3.2 \pm 0.2) \times 10^{12}$ cm$^{-2}$ can be obtained.

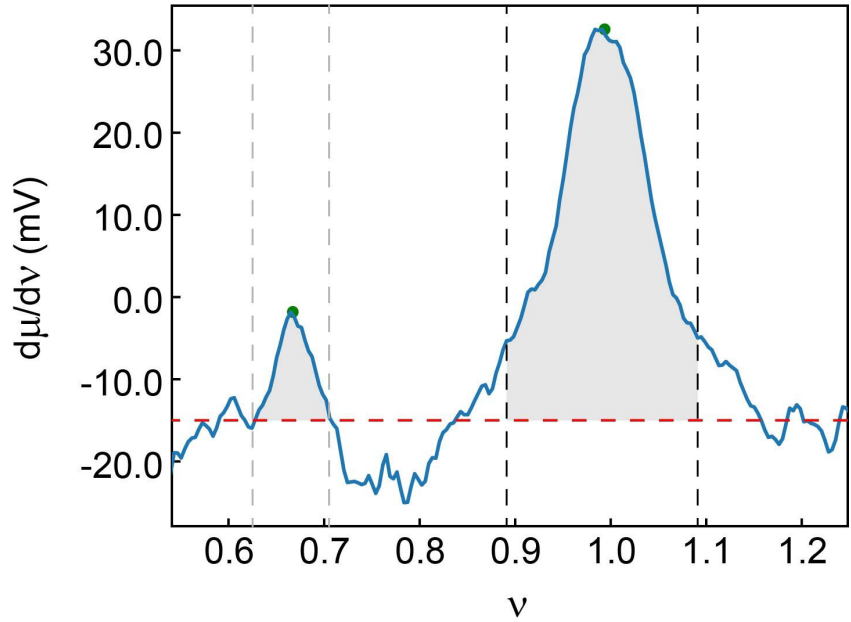

**Extended Data Figure 2 | Determination of the chemical potential jump and center-of-mass filling factors.** Filling factor dependent incompressibility at 1.6 K and zero electric and magnetic fields. We integrate the incompressibility peak above the base line (red dashed lines) to obtain the chemical potential jump at $\nu = 2/3$ and 1. This procedure effectively removes the negative incompressibility background in the chemical potential measurements. We also use the same incompressibility peak (above the background) to calculate the center-of-mass filling factors for the $\nu = 2/3$ and 1 states.

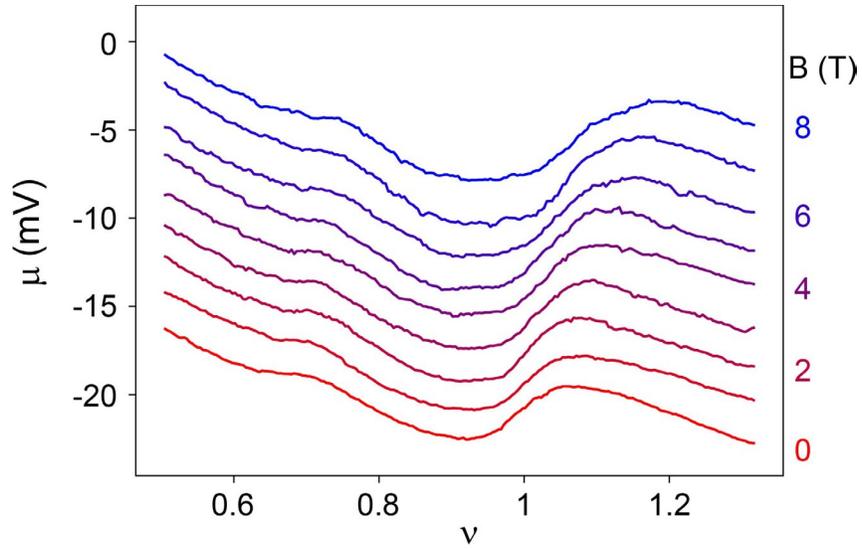

**Extended Data Figure 3 | Thermodynamic equation of state at varying magnetic fields.** The filling factor dependent chemical potential near zero electric field is shown. The curves are vertically displaced for clarity. We subtract the $B = 0$ T curve from the $B = 3$ T curve to obtain the $\left(\frac{\partial \mu}{\partial B}\right)_\nu$ plot in Fig. 2d of the main text.

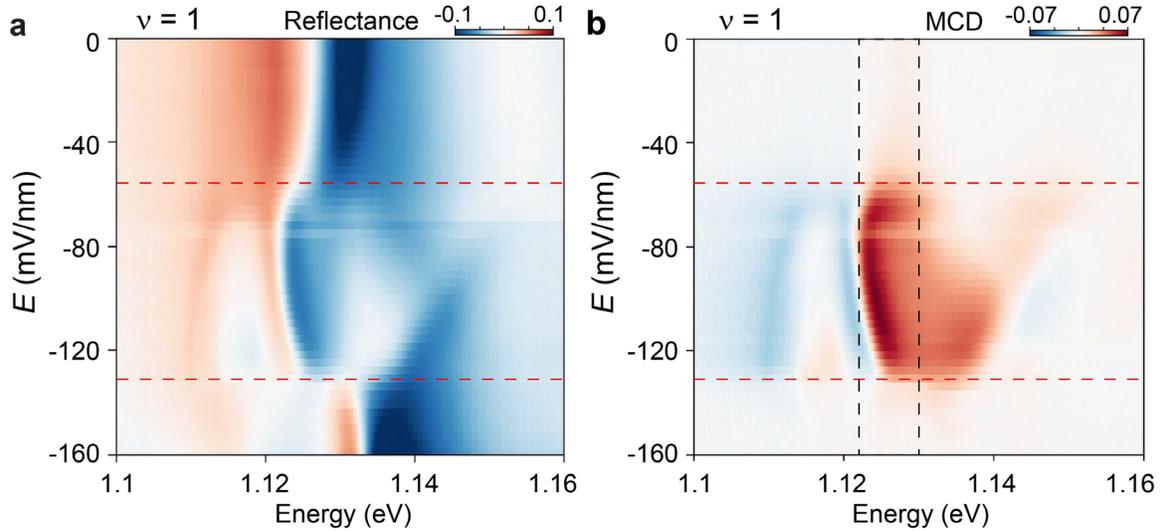

**Extended Data Figure 4 | Electric field dependent optical reflectance (a) and spontaneous MCD (b) spectra of twisted bilayer MoTe$_2$ at $\nu = 1$. a,** The depolarized spectra are shown. The bonding (~1.12 eV) and anti-bonding (~1.14 eV) features in the layer-hybridized region evolve into the neutral exciton feature (~1.13 eV) of one layer in the layer-polarized region. The red dashed lines denote the critical electric fields that separate the layer-hybridized and –polarized regions. We trace the highest absolute reflectance in order to obtain the 2D map in Fig. 3c in the main text. **b,** Strong spontaneous MCD is observed only in the layer-hybridized region. The black dashed lines bound the spectral window for the integration of MCD in order to obtain the 2D map in Fig. 3b in the main text.

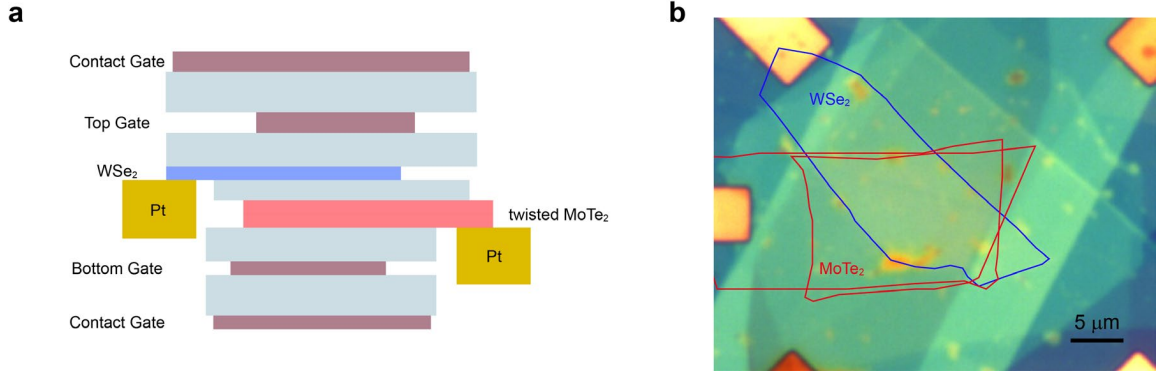

**Extended Data Figure 5 | Dual-gated twisted bilayer MoTe₂ devices for chemical potential measurements. a,** Schematic dual-gated device structure with a monolayer WSe₂ chemical potential sensor inserted in between the top gate and the twisted bilayer MoTe₂ sample. In addition to the structure shown in Fig. 1b in the main text, the contact gates to both the metal-sample and metal-sensor contacts are also shown. In our experiment, we apply a large negative voltage to the contact gates to heavily hole-dope the contact regions in order to achieve better electrical contacts for chemical potential measurements. **b,** Optical micrograph of the device shown in the main text. The red and blue lines outline the boundaries of the twisted bilayer MoTe₂ and the WSe₂ sensor, respectively.

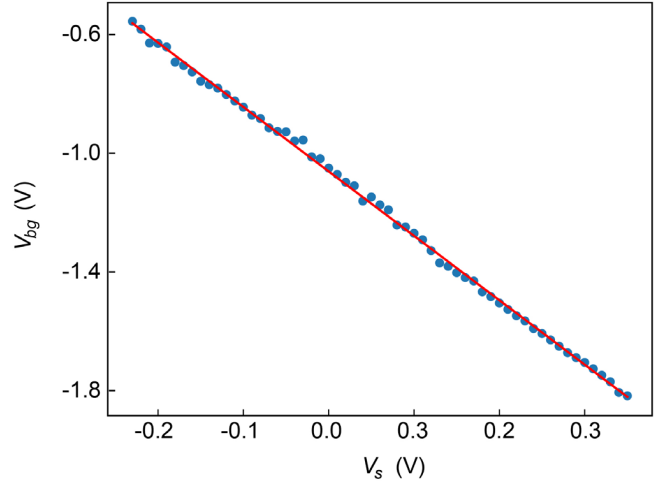

**Extended Data Figure 6 | Capacitance lever arm.** The center-of-mass filling factor for the $\nu = 1$ state versus the bottom gate voltage ($V_{bg}$) and the sample-sensor bias voltage ($V_s$). The slope determines the capacitance ratio $\frac{C_s}{C_{bg}} = 2.17 \pm 0.01$. Combined with the calibration of $C_s$ (Extended Data Fig. 1), the moiré density can be accurately determined.

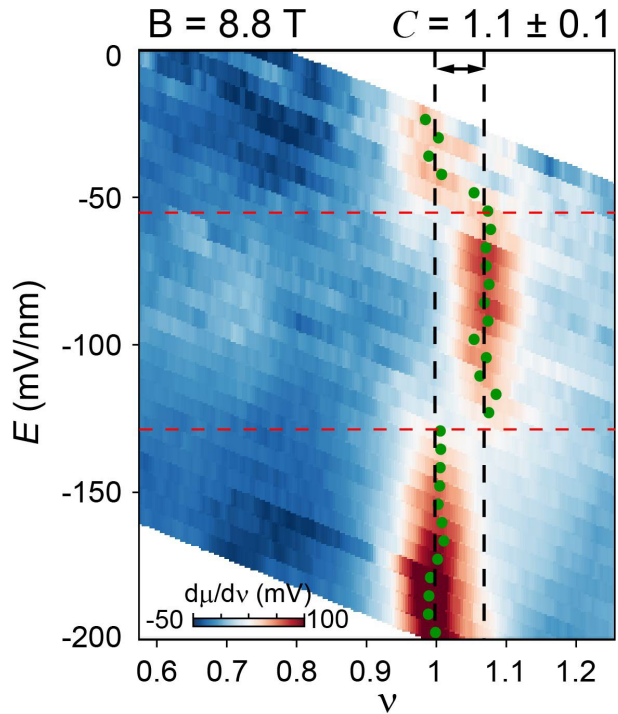

**Extended Data Figure 7 | Independent calibration of the $\nu = 1$ Chern number.**
Electric field and filling factor dependent incompressibility at 8.8 T and 4 K. A clear upshift in the filling factor for the $\nu = 1$ incompressible state is observed in the layer-hybridized region compared to the layer-polarized region (red dashed lines denote the boundaries). The upshift reflects the emergence of the Chern insulator (the correlated insulator in the layer-polarized region is non-topological). The average center-of-mass filling shift from the non-topological state (marked by the black dashed lines) is $0.070 \pm 0.006$ at 8.8 T. This corresponds to a Chern number $C = 1.1 \pm 0.1$ (see Methods). No non-topological insulating state at $\nu = 2/3$ can be identified.